\documentclass[epj]{svjour}
\usepackage{graphics}
\begin{document}
\newcommand{\eq}{\begin{eqnarray}}
\newcommand{\en}{\end{eqnarray}}
\title{Spectrum and decays of kaonic hydrogen}
\author{Ulf-G. Mei{\ss}ner\inst{1,2} \and Udit Raha\inst{1} \and 
Akaki Rusetsky\inst{1,3}
}                     
%
%
\institute{
Universit\"{a}t Bonn, Helmholtz-Institut f\"{u}r
Strahlen- und Kernphysik (Theorie),
Nu{\ss}allee 14-16, D-53115 Bonn, Germany 
\and
Forschungszentrum J\"{u}lich, Institut f\" {u}r Kenphysik
(Theorie), D-52425 J\"{u}lich, Germany
\and
On leave of absence from:
HEPI, Tbilisi State University, University st. 9, 380086 Tbilisi, Georgia
}
\date{Received: date / Revised version: date}
%
\abstract{
By using the non-relativistic effective
Lagrangian approach to bound states,
a complete expression for the isospin-breaking corrections to the energy
levels and the decay widths of kaonic hydrogen is obtained
up-to-and-including
$O(\alpha,m_d-m_u)$ in QCD. It is demonstrated that,
although the leading-order corrections at
$O(\alpha^{1/2},(m_d-m_u)^{1/2})$
emerging due to the unitarity cusp, are huge,
they can be expressed solely in terms of the $KN$ S-wave scattering
lengths.
Consequently, at leading order, it is possible to derive parameter-free
modified Deser-type relations, which can be used to extract the scattering
lengths from the hadronic atom data.
%
\PACS{
      {11.10.St}{}   \and
      {13.75.Jz}{}   \and
      {12.39.Fe}{}
     } 
} 
\maketitle

\section{Introduction}
\label{sec:intro}

The ongoing DEAR experiment~\cite{DEAR-proposal} at the DA$\Phi$NE
facility
(LNF-INFN) aims at an accurate measurement of the ground-state
strong energy shift and width of kaonic hydrogen, and the strong shift
of kaonic deuterium. Preliminary results of the measurements
for the kaonic hydrogen have been reported in Ref.~\cite{Cargnelli},
\eq\label{preliminary}
\Delta E_{1}^s= 202 \pm 45~ \mbox{eV}\, ,
\quad\quad \Gamma_{1} = 250 \pm 138~ \mbox{eV}\,.
\en
Here, $\Delta E_{1}^s$ stands for the strong-energy-level shift of the
ground
state of the kaonic hydrogen (total energy shift minus certain
electromagnetic
contributions), and $\Gamma_{1}$
denotes the width of the ground-state. It should be pointed out that these
results are in contradiction with the earlier
measurements~\cite{Davies,Izycki,Bird,KEK}, see also
Fig.~\ref{fig:plot} below.

The final goal of the DEAR experiment is to extract
precise values of the $KN$ S-wave scattering lengths from the data by
using
some counterpart of Deser-type relations \cite{Deser}. Neglecting isospin-breaking
corrections altogether,
in the case of kaonic hydrogen these relations are given by
\eq\label{Deser-type}
\Delta E_{1}^s-\frac{i}{2}\,\Gamma_{1}=-2\alpha^3\mu_c^2\, a_{K^-p}\, ,
\quad
a_{K^-p}=\frac{1}{2}\,(a_0+a_1)\, ,
\en
where $\mu_c$ denotes the reduced mass of the $K^-p$ system, and
$a_0,a_1$ stand for the $I=0,1$
S-wave $KN$ scattering lengths in QCD in the isospin
limit ($\alpha=0,~m_d=m_u$). In addition,
our definition of the isospin limit implies that
the particle masses in the multiplets
are taken equal to the charged particle masses in the real world
(proton, $\pi^+$, $K^+,\cdots$).
Further, in the experimental proposal it has been
stated that the precise knowledge of the $KN$ scattering lengths could
allow
one to deduce more accurate values for the $KN$ $\sigma$-terms and the
strangeness content of the nucleon. In practice, however, the
implementation
of the above program might pose a rather big challenge to theory (see,
e.g.
\cite{Gasser:DAFNE-Bonn} and references therein). For this reason, in this
paper we restrict
ourselves to the moderate goal of relating the $KN$ scattering lengths to
the
measurable characteristics of kaonic hydrogen at the accuracy that matches
the experimental precision. Using these scattering lengths for determining
the parameters of the low-energy kaon-nucleon interactions is thus out of
the
scope of the present paper.

It turns out that the isospin-breaking corrections to the
lowest-order relation given by Eq.~(\ref{Deser-type})
are huge. In particular, these are much larger than their
counterparts in pionic hydrogen, or in pionium (typically, a few percent).
This can be immediately seen, e.g. from the Table~\ref{tab:expansion} 
(see below) by comparing the entries in the same column.
The reason for this qualitative difference will be discussed.
In addition, the existing predictions in the literature, most of which are
done in the framework of potential scattering theory
(see~\cite{Dalitz,Deloff,all,Ivanov}
for an incomplete list of the earlier work on
the subject), are anything but
consistent with each other. In particular, although
large corrections have been predicted in some of these papers, these
effects
have not been treated systematically -- e.g., it is not always clear
whether
all possible large corrections are taken into account.
Needless to say, all this could make the interpretation of
the results of the accurate measurements of the DEAR experiment a
difficult
task.

The aim of the present paper is to obtain the formal relation between
the energy shift and the width of kaonic hydrogen, and the $KN$ scattering
lengths up-to-and-including isospin-breaking effects at
$O(\alpha,m_d-m_u)$\footnote{
We use throughout the Landau symbols $O(x)$ [$o(x)$] for quantities that
vanish like $x$ [faster than $x$] when $x$ tends to zero. Furthermore,
it is understood that this holds modulo logarithmic terms, i.e. we write
also
$O(x)$ for $x\ln x$.} in QCD, by using the systematic approach to the
bound-state problem based on the
non-relativistic effective Lagrangians.
In the past, this method has already been applied to study
pionium and kaonium~\cite{Labelle,the_others,pipi-piK}, as well as pionic
hydrogen
\cite{piN,piN1,Zemp}. It happens to be a very useful and convenient
approach to describe the spectrum and decays of this sort of
bound states. We shall see that the approach is
universal: the treatment of kaonic hydrogen closely follows the pattern of
pionic hydrogen.

\section{Formalism}

As it was already pointed out, the quantities that we are aiming to
extract from
the data on kaonic hydrogen, are the S-wave $KN$ scattering lengths
$a_0,~a_1$
evaluated in QCD in the isospin limit, i.e. in the absence of the
electromagnetic
interactions and at $m_d=m_u$.
Note that we avoid using the threshold scattering amplitude calculated
with
physical hadron masses but in the absence of electromagnetic effects,
which is sometimes encountered in the literature (see
e.g.,~\cite{Oset,Oller}).
The reason for this is that this quantity can not be consistently made
ultraviolet-finite to
all orders in Chiral Perturbation Theory (ChPT)\footnote{
Consider the loop contributions to the scattering amplitude
with both charged and neutral particles running inside the diagrams. 
The divergent parts which are generated by these loops, depend on charged and 
the neutral particle masses. Since the mass difference contains the 
electromagnetic piece proportional to $e^2$, in order to cancel all 
divergences one needs, along with the ``strong'' counterterm Lagrangian,
the ``electromagnetic'' counterterms as well.  
The latter was, however, ruled out from the beginning.}. 
Further, the isospin-breaking effects are
parameterized in terms of $\alpha$ and $m_d-m_u$. It is convenient to
introduce
a correlated counting of these effects, defining a formal parameter
$\delta\sim\alpha\sim m_d-m_u$~\cite{piN}. The Eq.~(\ref{Deser-type}) is
then
valid up-to-and-including $O(\delta^3)$ in isospin breaking, and to all
orders
in the chiral expansion for the quantity $a_0+a_1$ which is present in
this expression.
In the present paper we modify the relation (\ref{Deser-type}),
including all terms of order $\delta^{7/2}$ and $\delta^4$.

In order to construct a non-relativistic Lagrangian that can describe the
spectrum of kaonic hydrogen at $O(\delta^4)$, we note that:
\begin{itemize}
\item[i)]
The only states that are degenerate in mass with the $K^-p$ state in the
isospin
limit $\delta\to 0$, are the states $K^-p+n\gamma$, $\bar K^0n+n'\gamma$,
with
$n,n'=0,1,\cdots$. We explicitly ``resolve'' only these states in our
non-relativistic theory,
whereas the effect of other intermediate states, whose mass is not
degenerate with
that of the $K^-p$ state in the isospin limit, is included in the
couplings
of the non-relativistic effective Lagrangian.\footnote{For the
treatment of the $n\gamma$ intermediate
state in pionic hydrogen, which is similar in spirit to the approach used
here,
see~\cite{Zemp}.} In particular, the $SU(3)$
breaking scale $m_s-m_u$ counts at $O(\delta^0)$ in our approach.
As a result, all effects which are
non-analytic in the parameter $\delta$ -- e.g. containing $\sqrt{\delta}$
or
$\ln\delta$, should be produced by the loop expansion in the
non-relativistic theory.
To the contrary, the generic couplings $g_i$ of the non-relativistic
Lagrangian are
regular functions of $\delta$ and can be expanded in Taylor series
\eq\label{d_i}
g_i=g_i^{(0)}+\alpha g_i^{(1)}+(m_d-m_u)g_i^{(2)}+O(\delta^2)\, .
\en
\item[ii)]
The couplings $\tilde d_i$ that describe the $KN$ scattering
in the tree approximation (see Eq.~(\ref{Lagr-ini}))
are complex. The imaginary parts of the $\tilde d_i$ can be related
through
the unitarity condition to the transition cross sections of the
$KN$ initial state into the
different inelastic channels. In this case, there exist
open strong channels -- e.g. $\pi\Sigma,~\pi^0\Lambda$, etc.
The mass gap between these shielded two-particle
states and the $KN$ state is determined by the $SU(3)$ breaking scale.
Consequently, the couplings $\tilde d_i$ are complex already at
$O(\delta^0)$.
This is different from the case of the pionic hydrogen, where the
imaginary part
of the effective 2-pion--2-nucleon couplings is of order $\delta$.

\item[iii)]
The leading strong decay channel in the case of pionic hydrogen is
$\pi^0n$.
The phase space for this decay channel is proportional to
$(m_p+M_{\pi^+}-m_n-M_{\pi^0})^{1/2}$ and is thus suppressed by a
factor $\delta^{1/2}$. For this reason, the ratio of the decay widths into
the
leading electromagnetic channel $n\gamma$, and into the $\pi^0n$ channel
counts
as order $\delta^{1/2}$ only. Numerically, the branching ratio into the
$n\gamma$ channel amounts up to $\sim 40~\%$ in the total decay width. In
contrast
to this, in the case of kaonic hydrogen this branching ratio counts as
$O(\delta)$. The measured branching ratio into the leading
$\Lambda\gamma$, $\Sigma\gamma$
channels is much less than $1~\%$~\cite{Whitehouse} (the theoretical
description of this quantity by using chiral Lagrangians
\cite{Ivanov,Oset_Lee} gives the result consistent with the experiment
by order of magnitude). Consequently, the perturbative
treatment of the effects due to these channels, as it is carried out in
this paper,
is justified.\footnote{We thank E. Oset for interesting discussions on this 
issue.}

\item[iv)]
Within our approach, it is sufficient to deal with the sub-threshold
$\Lambda\,(1405)$
resonance indirectly, through the (large) $KN$ scattering lengths. The
reason
for this simplification is that the mass gap counts at $O(\delta^0)$ in
our
counting of the
isospin-breaking effects, i.e. the effect occurs at a ``hard scale'' and
should be included in the effective couplings.

\item[v)]
The $\bar K^0n$ system is a bit heavier than $K^-p$. This simple fact has
dramatic
consequences on the size of isospin-breaking corrections, which is nothing
but
the well-known cusp effect (note that the
cusp effect is also the dominant isospin-breaking effect in some other
low-energy processes, e.g. in neutral pion photo-production
off nucleons~\cite{cusp}.).
Namely, the loop with the $\bar K^0n$
intermediate state at threshold in the non-relativistic theory is
proportional
to $(m_n+M_{\bar K^0}-m_p-M_{K^+})^{1/2}\sim\sqrt{\delta}$ and is {\em
real}.
Its counterpart for the pionic hydrogen case
is {\em purely imaginary}. This means that,
in the case
of pionic hydrogen, the corrections to the {\em real} quantities -- the
energy shift and width -- can not contain a contribution from a single
neutral
loop. Only the product of two loops, which is a real quantity, can
contribute --
therefore, the corrections start at order $\delta$ with respect to the
leading-order
term. To the contrary,  
the corrections for kaonic hydrogen can contain a single neutral loop.
Due to this, the isospin-breaking corrections
to the Deser formula for kaonic hydrogen start at $O(\sqrt{\delta})$ and
are
much larger than their counterparts for pionic hydrogen.

\end{itemize}

Despite the differences between pionic and kaonic hydrogen that
were discussed above, one may apply exactly the same formal approach in
both cases
to calculate the bound state spectra. Below, we closely follow the path
outlined in Ref.~\cite{piN}. The effective non-relativistic Lagrangian is
given by
\eq\label{Lagr-ini}
{\mathcal L}&\,=\,&-\frac{1}{4}\, F_{\mu\nu}F^{\mu\nu}
+\psi^\dagger\,\biggl\{i{\mathcal D}_t-m_p+\frac{{\mathcal D}^2}{2m_p}
+\frac{{\mathcal D}^4}{8m_p^3}+\cdots
\nonumber\\[2mm]
&-&c_p^F\,\frac{e\sigma{\bf B}}{2m_p}
-c_p^D\,\frac{e({\mathcal D}{\bf E}-{\bf E}{\mathcal D})}{8m_p^2}
\nonumber\\[2mm]
&-&c_p^S\,\frac{ie\sigma({\mathcal D}\times{\bf E}-{\bf E}\times{\mathcal
D})}{8m_p^2}
+\cdots\biggr\}\,\psi
\nonumber\\[2mm]
&+&\chi^\dagger\,\biggl\{i\partial_t-m_n+\frac{\nabla^2}{2m_n}
+\frac{\nabla^4}{8m_n^3}+\cdots\biggr\}\,\chi
\nonumber\\[2mm]
&+&\sum_\pm (K^\pm)^\dagger\,\biggl\{iD_t-M_{K^+}
+\frac{{\bf D}^2}{2M_{K^+}}+\frac{{\bf D}^4}{8M_{K^+}^3}+\cdots
\nonumber\\[2mm]
&\mp& c^R_K\,\frac{e({\bf D}{\bf E}-{\bf E}{\bf
D})}{6M_{K^+}^2}
+\cdots\biggr\}\,K^\pm
\nonumber\\[2mm]
&+&{(\bar K^0)}^\dagger\,\biggl\{i\partial_t
-M_{\bar  K^0}+\frac{\nabla^2}{2M_{\bar K^0}}
+\frac{\nabla^4}{8M_{\bar K^0}^3}+\cdots\biggr\}\,\bar K^0
\nonumber\\[2mm]
&+&\tilde d_1\,\psi^\dagger\psi\,(K^-)^\dagger K^-+
\tilde d_2(\psi^\dagger\chi\,(K^-)^\dagger\bar K^0+h.c.)
\nonumber\\[2mm]
&+&\tilde d_3\chi^\dagger\chi\,(\bar K^0)^\dagger\bar K^0+\cdots\, .
\en
Here, $F_{\mu\nu}$ stands for the electromagnetic field strength tensor
(we work in the Coulomb gauge). Further,
$\psi$, $\chi$, $K^{\pm}$ and $\bar K^0$ denote the non-relativistic field
operators for the proton, neutron, charged and neutral kaon fields,
and ${\mathcal D}_t \psi=\partial_t \psi-ieA_0 \psi$,  
${\mathcal  D}\psi= \nabla \psi+ie{\bf A} \psi$,
$D_t K^\pm=\partial_t K^\pm\mp ieA_0 K^\pm$,
${\bf D} K^\pm= \nabla K^\pm\pm ie{\bf A} K^\pm$ are the covariant
derivatives acting on the proton and charged pion fields, respectively.
The ellipsis stand for the higher-dimensional operators and the UV
counterterms. The values of the couplings $c_i$ can be read off from
the matching condition for the kaon and nucleon electromagnetic
form-factors:
$c_p^F=1+\mu_p$,
$c_p^D=1+2\mu_p+\frac{4}{3}\, m_p^2\langle r_p^2\rangle$,
$c_p^S=1+2\mu_p$,
$c^R_K=M_{K^+}^2\langle r_K^2\rangle$,
where $\mu_p$ denotes the anomalous magnetic moment of the proton,
and $\langle r^2_p\rangle ,~\langle r^2_K\rangle $ stand for the
squared charge radii of the proton and the charged kaon, respectively.

\begin{figure}
\resizebox{0.35\textwidth}{!}{%
  \includegraphics{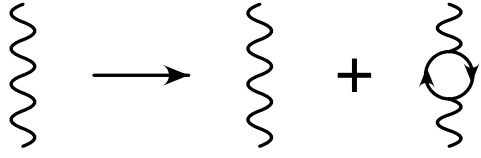}
}
\caption{Modification of the timelike component of the free photon propagator
  due to the electron vacuum polarization}
\label{fig:vacpol}
\end{figure}

Formally, the vacuum polarization contributions to the energy shift
and width start at $O(\alpha^5)$. It is, however, well known that the
electronic
vacuum polarization contribution is amplified by powers of the factor
$\mu_c/m_e\sim 10^3$, with $m_e$ being the electron mass.
It is convenient to count the quantity $\eta\doteq\alpha\mu_c/m_e$
as $O(\delta^0)$. Then, the leading-order contribution starts already at
$O(\delta^3)$, and the next-to-leading order contribution (interference of
strong
interactions with the vacuum polarization) comes at $O(\delta^4)$. In this
paper,
we explicitly include these contributions in the expression of the energy.
Note that at the accuracy
considered here, the whole effect of the vacuum polarization reduces to
the
modification of the timelike component of the free photon propagator by an
electron loop (see Fig.~\ref{fig:vacpol})

\eq\label{vac.pol}
D_{00}(k)&=&-\frac{1}{{\bf k}^2}\rightarrow
-\frac{1}{{\bf k}^2}
-\frac{\alpha}{3\pi}\,\int_{4m_e^2}^\infty
\frac{ds}{s+{\bf k}^2}\times
\nonumber\\[2mm]
&\times&\frac{1}{s}\,\biggl(1+\frac{2m_e^2}{s}\biggr)
\sqrt{1-\frac{4m_e^2}{s}}\, .
\en

\section{Energy shift and width of kaonic hydrogen}

The energy spectrum of kaonic hydrogen is obtained by using
Rayleigh-Schr\"o\-din\-ger perturbation theory (for details,
see Ref.~\cite{piN}, see also Ref.~\cite{Schweizer}
for the derivation of the
energy shift for a generic excited state of the $\pi K$ atom
by using non-relativistic
Lagrangians).
Namely, at the first step one constructs the
full Hamiltonian
${\bf H}={\bf H}_{\rm 0}+{\bf H}_{\rm C}+{\bf V}$
from the Lagrangian given by Eq.~(\ref{Lagr-ini}). In this expression,
${\bf H}_{\rm 0}$ stands for the free non-relativistic Hamiltonian of the
$KN$ pair, ${\bf H}_{\rm C}$ denotes the pure Coulomb interaction
between $K^-$ and the proton, and the rest of the interaction is included
in
the operator ${\bf V}$ which is treated as a perturbation.
The general solution of the
unperturbed Schr\"odinger equation with the Hamiltonian
${\bf H}_{\rm 0}+{\bf H}_{\rm C}$, that corresponds to the
quantum-mechanical
Coulomb problem for the bound system of spin-$0$ and spin-$1/2$ particles,
is characterized by the quantum numbers
$n=1,2,\cdots$, $j=\frac{1}{2},\frac{3}{2},\cdots$, $m=-j,\cdots j$ and
$l=j\pm \frac{1}{2}$,
\eq\label{Schrodinger}
&&({\bf H}_{\rm 0}+{\bf H}_{\rm C})|\Psi_{nljm}({\bf P})\rangle
=\bar E_n({\bf P})|\Psi_{nljm}({\bf P})\rangle\, ,
\nonumber\\[2mm]
&&\bar E_n({\bf P})=m_p+M_{K^+}+\frac{{\bf
P}^2}{2(m_p+M_{K^+})}-\frac{\mu_c\alpha^2}{2n^2}
\nonumber\\[2mm]
&&\hspace*{1.1cm}\doteq\bar E_n+\frac{{\bf
P}^2}{2(m_p+M_{K^+})}\, ,
\nonumber\\[2mm]
&&|\Psi_{nljm}({\bf P})\rangle=
\sum_s\int\frac{d^3{\bf q}}{(2\pi)^3}\,
\langle jm|l(m-s)\frac{1}{2}\, s\rangle \times
\nonumber\\[2mm]
&&\hspace*{1.77cm}\times Y_{l(m-s)}({\bf q})
\Psi_{nl}(|{\bf q}|)\,|{\bf P},{\bf q},s\rangle\, ,
\nonumber\\[2mm]
&&|{\bf P},{\bf q},s\rangle=
b^\dagger(\mu_1{\bf P}+{\bf q},s)a^\dagger(\mu_2{\bf P}-{\bf
q})|0\rangle\, .
\en
Here, $\mu_1=m_p/(m_p+M_{K^+})$, $\mu_2=M_{K^+}/(m_p+M_{K^+})$,
$a^\dagger$,
$b^\dagger$ stand for the creation operators for the non-relativistic
proton
and $K^-$,  $Y_{lm}$ and
$\langle jm|l(m-s)\frac{1}{2}\, s\rangle$ are the spherical harmonics and
the pertinent Clebsch-Gordan
coefficients, respectively, and $\Psi_{nl}(|{\bf q}|)$ denotes the radial
Coulomb wave function, which depends on the magnitude of the relative
momentum
$|{\bf q}|$. We further define the ``pole-removed''
Coulomb Green function $\hat {\bf G}_{nlj}(z)$ and the elastic transition
operator ${\bf M}_{nlj}(z)$
\eq\label{poleremoved}
&&{\bf G}_{\rm 0}(z)=\frac{1}{z-{\bf H}_{\rm 0}}\, ,\quad\quad
{\bf G}(z)=\frac{1}{z-{\bf H}_{\rm 0}-{\bf H}_{\rm C}}\, ,
\nonumber\\[2mm]
&&\hat {\bf G}_{nlj}(z)={\bf G}(z)-\sum_m\int\frac{d^3{\bf P}}{(2\pi)^3}\,
\frac{|\Psi_{nljm}({\bf P})\rangle\langle\Psi_{nljm}({\bf P})|}
{z-\bar E_n({\bf P})}\, ,
\nonumber\\[2mm]
&&{\bf M}_{nlj}(z)={\bf V}+{\bf V}\hat {\bf G}_{nlj}(z){\bf M}_{nlj}(z)\,
.
\en
By using Feshbach's formalism~\cite{Feshbach}, one can
show that the (complex) energy shift of the  level characterized
by the quantum numbers $njl$ is given by
\eq\label{master}
\Delta E_{nlj}&=&(\Psi_{nljm}|{\bf m}_{nlj}(\bar
E_n)|\Psi_{nljm})+o(\delta^4)
\, ,
\nonumber\\[2mm]
&&\langle\Psi_{nljm}({\bf P})|{\bf M}_{nlj}(z)|\Psi_{nljm}({\bf 0})\rangle
\nonumber\\[2mm]
&\doteq& (2\pi)^3\delta^3({\bf P})\,
(\Psi_{nljm}|{\bf m}_{nlj}(z)|\Psi_{nljm})\, .
\en

\begin{figure}
\resizebox{0.45\textwidth}{!}{%
  \includegraphics{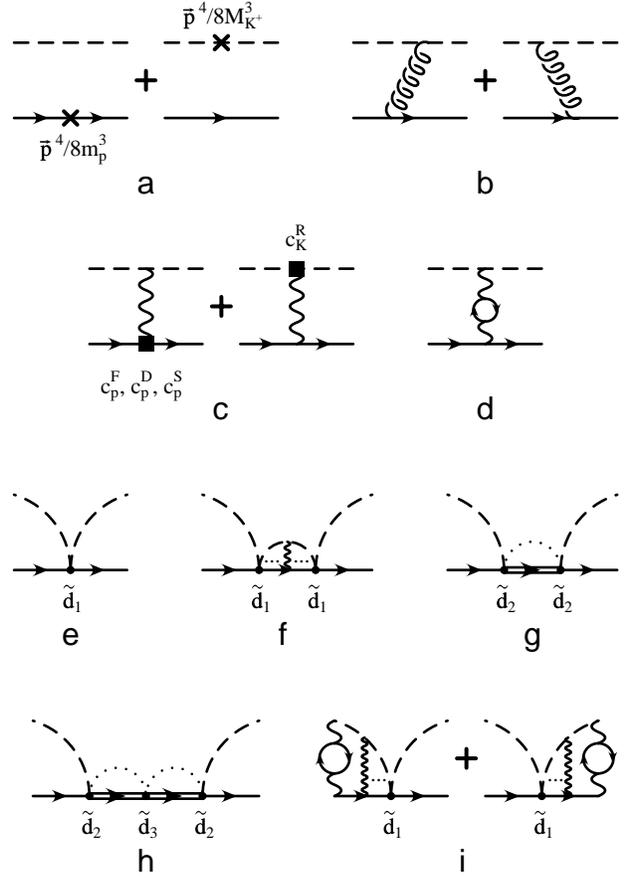}
}
\caption{The set of diagrams contributing to the energy shift of the
kaonic
hydrogen up-to-and-including $O(\delta^4)$. Solid, dashed, double, dotted,
wiggly and spring lines correspond to the proton, $K^-$, neutron, $\bar
K^0$,
Coulomb and transverse photons, respectively. The electrons run in the
closed
loops shown in diagrams (d) and (i). The diagrams (f) and (i) contain
Coulomb
ladders -- the contributions with $0,1,2,\cdots$ Coulomb photons
exchanged. }
\label{fig:diagrams} 
\end{figure}

In order to evaluate the energy shift up-to-and-in\-clu\-ding $O(\delta^4)$,
it
suffices to include only the diagrams $(a)-(i)$ shown in
Fig. \ref{fig:diagrams} in the
calculation of the operator ${\bf M}_{nlj}(z)$. The resulting expression
is then sandwiched between Co\-u\-lomb wave functions. We stress that none
of the other diagrams that can be constructed from the Lagrangian
(\ref{Lagr-ini}),
neither any possible contribution from higher-dimensional operators
that are not explicitly displayed in Eq.~(\ref{Lagr-ini}), contributes to
the energy shift at order $\delta^4$.
Here we do not provide the
details of the calculations, only the final result is given.
It is
convenient to introduce a well-defined splitting
of the energy shift into the ``electromagnetic'' and
``strong'' parts,\footnote{This naming scheme
should not be understood literally. For example, the ``electromagnetic''
contribution depends on the electromagnetic radii of the proton and $K^-$,
which are determined mainly by strong interactions. On the other hand,
there
are electromagnetic corrections to the couplings $\tilde d_i$.}
where the former does not contain the couplings
$\tilde d_i$
\eq\label{splitting}
\Delta E_{nlj}&=&\Delta E_{nlj}^{em}
+\delta_{l0}(\Delta E_n^s-\frac{i}{2}\,\Gamma_n)
+o(\delta^4)\, ,
\nonumber\\[2mm]
\Delta E_{nlj}^{em}&=&
-\frac{m_p^3+M_{K^+}^3}{8m_p^3M_{K^+}^3}\,\biggl(\frac{\alpha\mu_c}{n}\biggr)^4
\biggl\{\frac{4n}{l+\frac{1}{2}}-3\biggr\}
\nonumber\\[2mm]
&-&\frac{\alpha^4\mu_c^3}{4m_pM_{K^+}n^4}\,
\biggl\{-4n\delta_{l0}-4+\frac{6n}{l+\frac{1}{2}}\biggr\}
\nonumber\\[2mm]
&+&\frac{2\alpha^4\mu_c^3}{n^4}\,
\biggl(\frac{c_p^F}{m_pM_{K^+}}+\frac{c_p^S}{2m_p^2}\biggr)\,
\biggl\{\frac{n}{2l+1}-\frac{n}{2j+1}
\nonumber\\[2mm]
&-&\frac{n}{2}\,\delta_{l0}\biggr\}
+\frac{4\alpha^4\mu_c^3}{n^3}\,\delta_{l0}\biggl(\frac{c_p^D}{8m_p^2}
+\frac{c_K^R}{6M_{K^+}^2}\biggr)+\Delta E_{nl}^{(d)}\, ,
\nonumber\\[2mm]
\Delta E_n^s&-&\frac{i}{2}\,\Gamma_n\,=\,
-\frac{\alpha^3\mu_c^3}{\pi n^3}\,\biggl\{\tilde d_1
-\frac{\alpha\mu_c^2}{2\pi}\,\tilde d_1^{~2}\,(\chi+s_n(\alpha))
\nonumber\\[2mm]
&-&\tilde d_2^{~2}\,\frac{\mu_0q_0}{2\pi}
+\tilde d_2^{~2}\tilde d_3\,\biggl(\frac{\mu_0q_0}{2\pi}\biggr)^2\biggr\}
+\Delta E_n^{(i)}\, ,
\nonumber\\[2mm]
&&\hspace*{-1.2cm}\chi=\mu^{2(d-4)}\biggl(\frac{1}{d-4}-\Gamma'(1)-\ln 4\pi\biggr)
+\ln\frac{(2\mu_c)^2}{\mu^2}-1\, ,
\nonumber\\[2mm]
s_n(\alpha)&=&2(\psi(n)-\psi(1)-\frac{1}{n}+\ln\alpha-\ln n)\, ,
\en
and $\mu_0=m_nM_{\bar K^0}/(m_n+M_{\bar K^0})$,
$q_0=(2\mu_0(m_n+M_{\bar K^0}-m_p-M_{K^+}))^{1/2}$, 
$\psi(x)\doteq \Gamma'(x)/\Gamma(x)$.
Throughout the paper, we use dimensional regularization to tame
both UV and IR divergences. In the above formulae, $d$ stands for the
number of
space-time dimensions and $\mu$ denotes the scale of the dimensional
regularization. As a check on our calculations, we have verified that
the electromagnetic contributions in the above formulae that come from
the diagrams $(a)+(b)+(c)$, reproduce
the known result of Ref.~\cite{Austen} in the limit
$\langle r_p^2\rangle=\langle r_K^2\rangle=0$. Moreover, with
$c_p^F=c_p^S=0$
the above expressions reproduce the energy shift in the bound state of
two spin-$0$ particles~\cite{Schweizer}.

\begin{sloppypar}
The contributions due to the vacuum polarization
$\Delta E_{nl}^{(d)},~\Delta E_{n}^{(i)}$
have been evaluated e.g. in Ref.~\cite{Eiras}, in the same approach as
used
in the present paper (see also~\cite{Labelle,Jentschura} for the related
discussion).
At the order of accuracy we are working, these
contributions do not depend on the total angular momentum $j$. We do not
display here the (quite voluminous) general result for any $n$ and $l$.
The expression for $\Delta E_{nl}^{(d)}$ is given in Eq.~(3) of
Ref.~\cite{Eiras}. Furthermore, one may write $\Delta E_{n}^{(i)}=
-(\alpha^3\mu_c^3/\pi n^3)\,\tilde d_1\,
\delta_n^{\rm vac}+o(\delta^4)$,
and relate this quantity to the correction to the
bound-state wave function at the origin due to the vacuum polarization
effect
$\delta_n^{\rm vac}=2\,\delta\Psi_n(0)/\Psi_n(0)$. In Ref.~\cite{Eiras},
this
correction has been explicitly evaluated for the ground state (see Eq.~(6)
and Table II of this paper), although the method used in this paper
enables
one to make calculations for any excited level.
Here, it is important to stress that the ``electromagnetic''
contributions from diagrams $(a)+(b)+(c)+(d)$, which have to be
unambiguously identified and systematically evaluated up-to-and-including
$O(\delta^4)$, are only used for determining the so-called ``strong
shift''
(see Eq.~(\ref{Deser-type})) from which the information about the
$KN$ scattering lengths is extracted. Namely,
the strong shift is defined as a difference between the total
energy shift and the electromagnetic shift.
In the rest of the paper we deal with the strong shift only.
\end{sloppypar}

The equations (\ref{splitting}) do not solve our problem
completely: the energy shift is expressed in terms of the effective
couplings
$\tilde d_i$ which have still to be related to the observable quantities.
As in Refs.~\cite{pipi-piK,piN,Zemp,Schweizer}, this goal is achieved by
performing the matching for the $KN$ scattering amplitudes in the vicinity
of threshold. In the absence of isospin breaking one immediately gets
\eq\label{purelystrong}
\hspace*{-.3cm}
\tilde d_1^{~(0)}=\tilde d_3^{~(0)}=\frac{\pi}{\mu_c}\,(a_0+a_1)\,
,\quad
\tilde d_2^{~(0)}=\frac{\pi}{\mu_c}\,(a_1-a_0)\, ,
\en
where $\tilde d_i=\tilde d_i^{~(0)}+O(\delta)$. However, as one sees from
Eq.~(\ref{splitting}), in order to evaluate $\Delta E_{nlj}$
at $O(\delta^4)$, $\tilde d_1$ should be known at $O(\delta)$ (for
$\tilde d_2,~\tilde d_3$ the accuracy of Eq.~(\ref{purelystrong})
suffices).
At the required precision, the quantity $\tilde d_1$ can be determined
from
matching to the $K^-p$ threshold elastic scattering amplitude in the
presence of electromagnetic and strong isospin-breaking effects at
$O(\delta)$
-- the corresponding procedure is described in detail in
Refs.~\cite{piN,piN1}.
At the first step, one removes the one-photon exchange from the
spin-nonflip
part of the relativistic scattering amplitude for the process
$p(p)+K^-(q)\to p(p')+K^-(q')$
\eq\label{epjc}
&&\hspace*{-.4cm}T_{KN}=\bar u(p')\biggl\{ \tilde D(s,t)
-\frac{1}{4m_p}\,[\not\! q',\not\! q]\,\tilde B(s,t)\biggr\} u(p)\, ,
\nonumber\\[2mm]
&&\hspace*{-.4cm}\tilde D'(s,t)=\tilde D(s,t)-e^2F_K(t)F_1(t)(s-u)/(2m_pt)\, ,
\en
where $F_K(t)$, $F_1(t)$ denote the kaon electromagnetic and the
nucleon Dirac formfactors, respectively, and $s,t,u$ are the usual
Mandelstam
variables. The quantity $\tilde D'(s,t)$ is singular at threshold,
as the magnitude of the
relative 3-momentum of the proton and kaon in the CM frame $|{\bf p}|$
vanishes. At $O(\delta)$, the structure of this singularity
is given by~\cite{piN,piN1}
\eq\label{threshold}
{\rm e}^{-2i\alpha\theta_C(|{\bf p}|)}\,\tilde D'(s,t)\biggr|_{|{\bf
p}|\to 0}
&=&\frac{\tilde B_1}{|{\bf p}|}+\tilde B_2\ln\frac{|{\bf p}|}{\mu_c}
+{\mathcal T}_{KN}
\nonumber\\[2mm]
&+&O(|{\bf p}|)\, ,
\en
where $\theta_C(|{\bf p}|)$ denotes the (dimensionally regularized
in\-fra\-red-di\-ver\-gent) Cou\-lomb phase
\eq\label{phase}
\theta_C(|{\bf p}|)&=&\frac{\mu_c}{|{\bf p}|}\,\mu^{d-4}\biggl(\frac{1}{d-4}
-\frac{1}{2}\,(\Gamma'(1)+\ln 4\pi)
\nonumber\\[2mm]
&+&\ln\frac{2|{\bf p}|}{\mu}\biggr)\, .
\en
In this normalization, the S-wave $KN$ scattering lengths and the
threshold
amplitude ${\mathcal T}_{KN}$ in the isospin limit are related by
\eq\label{sc_l}
{\mathcal T}_{KN}=4\pi\biggl(1+\frac{M_{K^+}}{m_p}\biggr)
\,\frac{1}{2}\,(a_0+a_1)+O(\sqrt{\delta})\, .
\en
The quantity ${\mathcal T}_{KN}$ should be matched to its non-re\-la\-ti\-vis\-tic
counterpart ${\mathcal T}_{KN}^{NR}$, written in terms of the couplings
$\tilde d_i$. A direct calculation with the Lagrangian
Eq.~(\ref{Lagr-ini}),
which is carried out in a similar way as in Ref.~\cite{piN}, yields
\eq\label{TKN_NR}
{\mathcal T}_{KN}^{NR}&=&
\tilde d_1-\tilde d_2^{~2}\,\frac{\mu_0q_0}{2\pi}
+\tilde d_2^{~2}\tilde d_3\,\biggl(\frac{\mu_0q_0}{2\pi}\biggr)^2
\nonumber\\[2mm]
&-&\tilde d_1^{~2}\,\frac{\alpha\mu_c^2}{2\pi}\,(\chi-2\pi i)\, .
\en
The matching condition $2M_{K^+}{\mathcal T}_{KN}^{NR}={\mathcal T}_{KN}$
enables
one to determine the coupling $\tilde d_1$ at the required accuracy.
Substituting this value of $\tilde d_1$ into the expression
for the strong shift, we finally
get the formula in terms of the observable quantities,
which contains all isospin breaking terms up-to-and-including
$O(\delta^4)$
\eq\label{general_n}
\Delta E_n^s-\frac{i}{2}\,\Gamma_n&=&-\frac{\alpha^3\mu_c^3}{2\pi
M_{K^+}n^3}\,
{\mathcal T}_{KN}\biggl\{1
-\frac{\alpha\mu_c^2}{4\pi M_{K^+}}\times
\nonumber\\[2mm]
&\times&{\mathcal T}_{KN}(s_n(\alpha)+2\pi i)
+\delta_n^{\rm vac}\biggr\} .
\en
Although Eq.~(\ref{general_n}) formally solves the problem
stated in the introduction, it is still not well suited for the analysis
of the experimental data. The reason for this is clear from
Eq.~(\ref{TKN_NR}). There it is immediately seen
that the unitarity correction
from the $\bar K^0 n$ bubble (second term in the r.h.s. of this equation),
whose counterpart in the bound-state sector is depicted in
Fig.~\ref{fig:diagrams}g, starts to contribute to the isospin-breaking
part of ${\mathcal T}_{KN}^{NR}$ and ${\mathcal T}_{KN}$  at
$O(\sqrt{\delta})$ (the quantity
$q_0$ is of
order $\sqrt{\delta}$). The situation here differs from the pionic
hydrogen case where the counterpart of the quantity $q_0$ is imaginary
because
$m_p+M_{\pi^+}>m_n+M_{\pi^0}$, and the imaginary part of ${\mathcal
T}_{\pi N}$
starts at $O(\sqrt{\delta})$, not $O(1)$. Consequently, in the $\pi N$
case
the analog of Eq.~(\ref{sc_l}) for the real part of the scattering
amplitude
contains corrections at order $\delta$ and not $\sqrt{\delta}$. Exactly
the
above-described effect, which is nothing but the unitarity cusp in the
$K^-p$ elastic amplitude, is the source of the huge isospin-breaking
corrections in the energy shift of kaonic hydrogen, which were mentioned
in the
introduction.

The above problem can be solved in the following manner. From
Eq.~(\ref{TKN_NR}) we see that only the unitarity correction 
to the quantity ${\mathcal T}_{KN}^{NR}$
behaves like
$\sqrt{\delta}$, and all other corrections, including corrections in the
couplings $\tilde d_i$, start at $O(\delta)$ and are regular in $\delta$ 
at this order.
On the other hand, from Eq.~(\ref{purelystrong}) it is seen that the
quantity
$\tilde d_2$ can be written in terms of the scattering lengths, up to the
terms of order $\delta$. From this one concludes that the corrections
at $O(\sqrt{\delta})$, albeit big, are expressed only in terms of the same
scattering lengths, that are already present in the Deser-type relations
at the
leading order,\footnote{The same is true for the non-analytic corrections
from Eq.~(\ref{general_n}), which are proportional to $\alpha\ln\alpha$.}
and the structure-dependent corrections start at $O(\delta)$.
This counting can be implemented in the Deser-formula, e.g. as follows.
We sum up any number of strong neutral bubbles shown in
Fig.~\ref{fig:diagrams}g, since the first term in this expansion contains
exactly the desired singular piece with $\sqrt{\delta}$. Further,
instead of Eq.~(\ref{sc_l}), we write
\eq\label{sc_lnew}
{\mathcal T}_{KN}&=&{\mathcal T}_{KN}^{(0)}
+\frac{i\alpha\mu_c^2}{2M_{K^+}}\,({\mathcal T}_{KN}^{(0)})^2
+\delta{\mathcal T}_{KN}+o(\delta)\, ,
\nonumber\\[2mm]
{\mathcal T}_{KN}^{(0)}&=&4\pi\biggl(1+\frac{M_{K^+}}{m_p}\biggr)
\frac{\frac{1}{2}\,(a_0+a_1)+q_0a_0a_1}{1+\frac{q_0}{2}\,(a_0+a_1)}\, .
\en
The above equation is nothing but the definition of $\delta{\mathcal
T}_{KN}$,
and our statement amounts to 
\eq\label{deltaT}
\delta{\mathcal T}_{KN}=\alpha\, t_1 +(m_d-m_u)\, t_2 +o(\delta)\, ,
\en
where $t_1,t_2$ are functions of $\hat m=\frac{1}{2}\,(m_u+m_d),~m_s$ and
$\Lambda_{QCD}$. For the
actual calculation of $\delta{\mathcal T}_{KN}$ one has, e.g.
as in the $\pi N$ case, to use the information about the underlying
dynamics
which is contained in the low-energy effective chiral meson-baryon
Lagrangians.
With the definition given in Eq.~(\ref{sc_lnew}), the formula for the
strong
shift looks similar to that for the case of the pionic hydrogen
\eq\label{final}
\Delta E_n^s-\frac{i}{2}\,\Gamma_n&=&-\frac{\alpha^3\mu_c^3}{2\pi
M_{K^+}n^3}\,
({\mathcal T}_{KN}^{(0)}+\delta{\mathcal T}_{KN})\times
\nonumber\\[2mm]
&\times&\biggl\{1
-\frac{\alpha\mu_c^2s_n(\alpha)}{4\pi M_{K^+}}\,{\mathcal T}_{KN}^{(0)}
+\delta_n^{\rm vac}\biggr\} ,
\en
where ${\mathcal T}_{KN}^{(0)}$, $\delta{\mathcal T}_{KN}$ are given by
Eq.~(\ref{sc_lnew}), and $\delta{\mathcal T}_{KN}=O(\delta)$.
This is the final formula, which is best suited for the analysis of the
experimental data.

\section{Results, discussion and higher order corrections}

The isospin-breaking corrections at $O(\sqrt{\delta})$ that
are contained in the relation of ${\mathcal T}_{KN}^{(0)}$ to the S-wave
$KN$ scattering lengths $a_0,a_1$, are numerically by far the dominant
ones. These corrections have been derived more
than 40 years ago~\cite{Dalitz,Deloff} by using the $K$-matrix formalism.
However, this piece of information should be
still supplemented by the arguments in favor of the conjecture
that the remaining corrections are small, i.e.
$\delta{\mathcal T}_{KN}=O(\delta)$, as done in the present paper.

In order to get a feeling how big the corrections to the Deser formula
can be, we have done a
simple exercise. In table \ref{tab:expansion} we list the results of the
expansion of the quantity ${\mathcal T}_{KN}^{(0)}$ in powers of
$q_0\sim\sqrt{\delta}$, so that ${\mathcal T}_m^{(0)}$ denotes
${\mathcal T}_{KN}^{(0)}$ calculated up-to-and-including $O(q_0^m)$
and ${\mathcal T}_{KN}^{(0)}\doteq{\mathcal T}_\infty^{(0)}$.
The values for the scattering lengths that are needed in these
calculations,
are taken from Ref.~\cite{Oller}.\footnote{The scattering lengths
$a_0,a_1$ are not displayed separately in Ref~\cite{Oller}. We thank J.
Oller
for providing these values.} In these papers, the
$KN$ scattering amplitudes are obtained by iterating tree-level
diagrams calculated within ChPT, through the Lippmann-Schwinger equation
(see also~\cite{Oset,KSW} for  earlier references).
In addition, we use the experimental values of the scattering lengths
given in Ref.~\cite{Martin}.
As we see from the table, the corrections at $O(\sqrt{\delta})$ are indeed
huge -- they amount up to a few tens of percent. More precise predictions
are
not possible because the scattering lengths themselves are not very well
known.
On the other hand, instead of $\frac{1}{2}\,(a_0+a_1)$ as contained
in the original Deser formula, one might determine the
combination of these scattering lengths  
${\mathcal T}_{KN}^{(0)}$.
This combination, which already includes the
big corrections at $O(\sqrt{\delta})$, can be extracted from data with a
much better accuracy.
As one may observe from Table~\ref{tab:expansion},
the convergence of the expansion in the parameter
$\sqrt{\delta}$ is rather good.
\begin{table}[t]
\begin{center}
\begin{tabular}{|l|l|l|l|}
\hline\hline
& Ref.~\cite{Oller} & Ref.~\cite{Martin} \\
& $a_0=-1.31+1.24i$ & $a_0=-1.70+0.68i$ \\
& $a_1=0.26+0.66i$  & $a_1=0.37+0.60i$  \\
\hline
${\mathcal N\,T}_0^{(0)}$ &$-0.52+ 0.95i$
& $-0.66+ 0.64i$ \\
${\mathcal N\,T}_1^{(0)}$ &$-0.68+ 1.09i$
& $-0.98+ 0.66i$ \\
${\mathcal N\,T}_2^{(0)}$ &$-0.67+ 1.15i$
& $-1.04+ 0.73i$ \\
${\mathcal N\,T}_3^{(0)}$ &$-0.65+ 1.16i$
& $-1.04+ 0.75i$ \\
\hline
${\mathcal N\,T}_\infty^{(0)}$ &$-0.65+
1.15i$  & $-1.03+ 0.76i$ \\
\hline\hline
\end{tabular}
\end{center}


\caption{Expansion of the $KN$ scattering
amplitude ${\mathcal T}_{KN}^{(0)}$ in powers of $q_0$. Scattering lengths
and amplitudes are given in fm, and 
${\mathcal N}\doteq \bigl(4\pi(1+M_{K^+}/m_p)\bigr)^{-1}$.\label{tab:expansion}}
\end{table}

Note also that if one uses the scattering lengths $a_0=-2.24+1.94i$,
$a_1=0.54 +0.54i$ (in fm) as given in Ref.~\cite{Oset}, then the convergence
of the series in $\sqrt{\delta}$ is significantly worse. More precisely,
one gets:
$-0.85+1.24i$, $-1.28+1.82i$, $-1.17+2.12i$, $-1.04+2.16i$,
$\cdots -1.00+2.09i$ (in the notation employed in
Table~\ref{tab:expansion}).
It can be argued that this result contradicts the general expectation
about the size of the isospin-breaking corrections: one sees that in these
series the corrections at $O(\delta^{3/2})$ amount approximately  to
$15-20~\%$
in the real part of the amplitude. Moreover, this result also contradicts 
our derivation of the energy shift: there is no justification for neglecting
the $O(\delta^{3/2})$ terms in the bound-state energy, if in the amplitude their contribution is so large. 
However, we would like to stress that
in the approach used in Ref.~\cite{Oset}, the straightforward introduction
of a cutoff to regularize the unitarity resummation violates chiral
symmetry,
since the amplitudes are not
matched to the chiral expansion\footnote{For a recent discussion on the
use of cutoff regularization
in chiral effective field theories, see \cite{BHM}.}.
The considerably larger values of the scattering
lengths than those determined by the experiment \cite{Martin},
which in its turn cause problems concerning the convergence of the series,
might have been resulted from this implicit violation of chiral symmetry.
Such a problem does not arise in the calculation of Ref.~\cite{Oller} since
an explicit matching to the ChPT amplitudes is performed and the
regularization
is done employing by subtracted dispersion relations. That this procedure
leads to reasonable scattering parameters was demonstrated
explicitly for the case of pion-nucleon scattering in \cite{OM}.

We further investigate the magnitude of $O(\delta)$ corrections in
Eq.~(\ref{final}). The Coulomb corrections that are amplified by
$\ln\alpha$,
are sizable but smaller than those
due to the unitarity cusp: for the choice of
scattering lengths from Refs.~\cite{Oller,Martin}, the real part of the
correction term in the ground state is $[9\%,~15\%]$, respectively.
Again, we do not need
to know this number very accurately; since the Coulomb correction
depends on the scattering lengths $a_0,a_1$ only,
we can use the modified Deser
relation which includes the Coulomb term to
determine ${\mathcal T}_{KN}^{(0)}$.

\begin{sloppypar}
Below, we briefly consider other corrections contained in
Eq.~(\ref{final}).
The calculation of the quantity $\delta{\mathcal T}_{KN}$ proceeds
analogously
to the $\pi N$ case. This quantity starts at order $p^2$ in ChPT. 
Further, according to Eq.~(\ref{sc_lnew}), it is equal to the isospin-breaking
part of ${\mathcal T}_{KN}$ at this order. In the
actual calculations we have used strong
(see, e.g.~\cite{Krause,Jenkins,BKM,Kaiser}) and electromagnetic
meson-baryon
Lagrangians of $SU(3) \times SU(3)$ ChPT (for the construction principles,
see
\cite{MMS})
\end{sloppypar}
\eq\label{krause}
{\mathcal L}_2&=&
b_0{\rm Tr}(\bar BB){\rm Tr}(\chi_+)+b_D{\rm Tr}(\bar B\{\chi_+,B\})
\nonumber\\[2mm]
&+&b_F{\rm Tr}(\bar B[\chi_+,B])
+F_0^2G_1{\rm Tr}(Q_+^2){\rm Tr}(\bar B B)
\nonumber\\[2mm]
&+&F_0^2G_2{\rm Tr}(Q_-^2){\rm Tr}(\bar B B)
+F_0^2G_3{\rm Tr}(\bar B Q_+^2B)
\nonumber\\[2mm]
&+&F_0^2G_4{\rm Tr}(\bar B Q_+BQ_+)
+F_0^2G_5{\rm Tr}(\bar B BQ_+^2)
\nonumber\\[2mm]
&+&F_0^2G_6{\rm Tr}(\bar B Q_-^2B)
+F_0^2G_7{\rm Tr}(\bar B Q_-BQ_-)
\nonumber\\[2mm]
&+&F_0^2G_8{\rm Tr}(\bar B BQ_-^2)
+\mbox{terms with derivatives}\, ,
\en
\noindent\begin{sloppypar}\noindent
where ${\mathcal M}={\rm diag}(m_u,m_d,m_s)$ and
$Q=e\,{\rm diag}(2/3,-1/3,-1/3)$ are the quark mass matrix and the charge
matrix, respectively,
$B$ stands for the baryon octet field,
$\chi_+=2B_0(u{\mathcal M}u+u^\dagger{\mathcal M}u^\dagger)$,
$Q_\pm=\frac{1}{2}\,(uQu^\dagger\pm u^\dagger Qu)$, and
$U=u^2={\rm exp}(\frac{i}{F_0}\,\Phi)$,  with
$\Phi$ being the pseudoscalar boson octet field.
Here, $F_0$ is the Goldstone boson decay constant (in the chiral limit)
and $B_0 = |\langle0|\bar q q|0\rangle|/F_0^2$ measures the strength
of the quark-antiquark condensate (in the chiral limit). 
Further, $b_0,b_D,b_f$ and $G_i$ denote $O(p^2)$ strong and
electromagnetic
low-energy constants (LECs), respectively.
Note that since our definition of the isospin limit involves the physical
masses of the charged particles, the terms with  derivatives of the kaon
fields do not contribute to the isospin-breaking part of the $K^-p$
elastic amplitude at $O(p^2)$.
This quantity is equal to
\end{sloppypar}
\eq\label{dT}
\delta{\mathcal T}_{KN}&=&\frac{4(\Delta M_K^2)_{\rm em}}{F_0^2}\,
(b_0+b_D)+e^2\biggl\{-G_1+G_2-\frac{2}{3}\,G_3
\nonumber\\[2mm]
&+&\frac{1}{2}\,G_4-\frac{1}{3}\, G_5+\frac{1}{2}\,G_6
+\frac{1}{2}\, G_8\biggr\}\, ,
\en
where the subscript `em' denotes the electromagnetic mass shift.
Using Da\-shen's theorem, we can replace $(\Delta M_K^2)_{\rm em}$
by $(\Delta M_\pi^2)_{\rm em}=\Delta M_\pi^2+\cdots$ at this order.
Further, in the
numerical estimates we use
$b_0=-0.517~{\rm GeV}^{-1}$, $b_D=0.066~{\rm GeV}^{-1}$ from
Ref.~\cite{BKM}.
For the unknown electromagnetic LECs the order-of-magnitude estimate
$e^2F_0^2|G_i|\leq \alpha m_p/2\pi$ was used. With these numerical values one
gets
$\delta{\mathcal T}_{KN}/{\mathcal T}_{KN}^{l.o.}=(-0.5\pm 0.4)\%$ at
$O(p^2)$,
where
${\mathcal T}_{KN}^{l.o.}=M_{K^+}/F_0^2$ is the isospin-symmetric part of
the $K^-p$
amplitude at the leading order. The fact that the uncertainty in the
isospin-breaking part of the $KN$ amplitude turns out to be smaller than
in
the $\pi N$ amplitude \cite{piN1} is not surprising: the quantities
$\delta {\mathcal T}_{KN}$ and $\delta {\mathcal T}_{\pi N}$ are of the
same
order of magnitude, whereas the isospin-symmetric part in the $KN$
amplitude
is increased by a factor $M_{K^+}/M_{\pi^+}$ and additionally by a
group-theoretical factor of $2$.
The results are anyway to be taken with a grain of salt: it needs to be
seen how the results are changed in higher orders in chiral expansion.
Finally, we note that the value $\delta_1^{\rm vac}=0.87\%$ given
in Ref.~\cite{Eiras} suggests that at present accuracy one may well
include this term in the systematic error.

\begin{figure}
\resizebox{0.45\textwidth}{!}{%
  \includegraphics{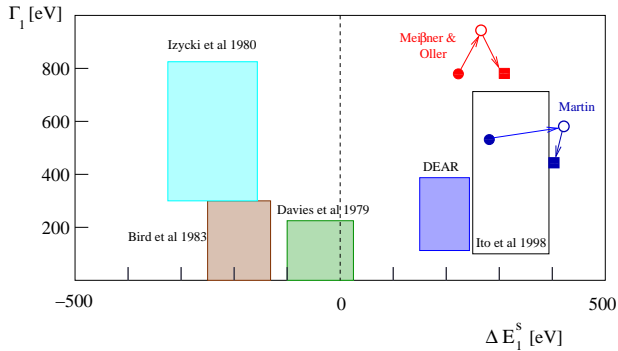}
}
\caption{Predictions of the ground-state strong shift $\Delta E_1^s$ and width
$\Gamma_1$.
Filled  circles correspond to using the Deser formula (\ref{Deser-type}),
empty circles to using ${\mathcal T}_{KN}^{(0)}$ instead of
$\frac{1}{2}\,(a_0+a_1)$ in this formula, and filled boxes to our final
formula (\ref{final}) with $\delta{\mathcal T}_{KN}=\delta_n^{\rm vac}=0$.}
\label{fig:plot}       
\end{figure}

In order to visualize the size of the isospin-breaking effects, in
Fig.~\ref{fig:plot} we plot the theoretical predictions corresponding to
the
scattering lengths from Refs. \cite{Oller,Martin} versus the old and
new experimental measurements of the energy spectrum of  kaonic hydrogen
\cite{Cargnelli,Davies,Izycki,Bird,KEK}. As we immediately observe from 
the plot, the use of the lowest-order Deser formula Eq.~(\ref{Deser-type})
can not be justified any more: both
cusp effect and Coulomb corrections have a size comparable with the
present
precision of the DEAR experiment, and should be taken into account during
extracting S-wave $KN$ scattering lengths from the experimental data.
Note an apparent discrepancy between the recent DEAR measurements and the
predictions obtained by using the scattering lengths from 
Refs.~\cite{Oller,Martin}. Moreover, one sees that
when the isospin-breaking corrections are applied, the results move away
from the DEAR measurements. We conclude, that the 
further investigations are needed in order to
shed light on this interesting issue.

\section{Summary and outlook}

In this paper we derived the formal expression for the strong
shifts of the energy levels in kaonic hydrogen in QCD, up-to-and-including
$O(\delta^4)$ in the isospin breaking parameter
$\delta\sim\alpha,m_d-m_u$.
The use of the non-relativistic effective Lagrangian approach allows one
to treat that otherwise extremely complicated problem with a surprising
ease.
We discover that large isospin-breaking corrections arise, in particular,
due to the following
sources: (a) $s$-channel rescattering with the $\bar K^0n$ intermediate
state
(cusp effect), and (b) Coulomb corrections that are non-analytic in
$\alpha$.
We further prove that the remaining corrections are analytic in $\delta$ at
$O(\delta)$.
Examining some of these corrections, on the other hand, we do not find a
big
effect -- the obtained values are at the percent level, which one expects
to be a typical size of isospin breaking in QCD.

The present status of corrections in kaonic hydrogen
can be summarized by the Eq.~(\ref{final}).
Instead of the combination $\frac{1}{2}\,(a_0+a_1)$
which enters in the original Deser formula (\ref{Deser-type}), we propose
to focus on the extraction of the quantity ${\mathcal T}_{KN}^{(0)}$ from
the experimental data. The reason for
this is that ${\mathcal T}_{KN}^{(0)}$ already includes the dominant
non-analytic
corrections in a parameter-free form. The remaining analytic corrections
at
$O(\delta)$ are contained in the quantities
$\delta {\mathcal T}_{KN}$ and $\delta_n^{\rm vac}$.
The evaluation of $\delta {\mathcal T}_{KN}$ within ChPT could be
interesting,
but possibly complicated due to the expansion in the strange quark mass.
At the present stage, in the absence of such calculations, the best is
to include
$\delta {\mathcal T}_{KN}$ in the estimate of the systematic error.
{}From the above discussion one may hope that the effect from
$\delta {\mathcal T}_{KN}$ should not exceed a few
percent, which is a natural size of electromagnetic corrections.


\section*{Acknowledgements}
The authors thank J. Gasser, A. Ivanov,
C. Petrascu, J. Schweizer and P. Zemp for useful discussions.
We are particularly grateful to J\"urg Gassser for a careful reading
of the manuscript.

\end{document}